\documentclass[a4paper,11pt]{JHEP}

\usepackage{graphicx}
\usepackage{epsfig}
\usepackage{amsmath, amssymb}
\usepackage{dcolumn}% Align table columns on decimal point
\usepackage{bm}% bold math
\usepackage{amsfonts}
\newcommand{\bx}{{\mathbf x}}

\newcommand{\RR}{{\cal R}}
\newcommand{\FF}{{\cal F}}

\newcommand{\al}{\alpha}

\newcommand{\de}{\delta}
\newcommand{\De}{\Delta}
\newcommand{\ep}{\epsilon}

\newcommand{\La}{\Lambda}
\newcommand{\la}{\lambda}
\newcommand{\Om}{\Omega}

\newcommand{\si}{\sigma}

\newcommand{\ra}{\rightarrow}

\newcommand{\be}{\begin{equation}}
\newcommand{\ee}{\end{equation}}

\newcommand{\bea}{\begin{eqnarray}}
\newcommand{\eea}{\end{eqnarray}}
\newcommand{\bean}{\begin{eqnarray*}}
\newcommand{\eean}{\end{eqnarray*}}
\newcommand{\dd}{\partial}

\title{Back Reaction from Walls}

\author{Enea Di Dio
\\D\'epartement de Physique Th\'eorique and Center for Astroparticle Physics,\\
Universit\'e de Gen\`eve,\\ 24 quai Ernest 
Ansermet, CH--1211 Gen\`eve 4, Switzerland\\
\email{enea.didio@unige.ch}
}
\author{Marc Vonlanthen
\\ D\'epartement de Physique Th\'eorique and Center for Astroparticle Physics,\\
Universit\'e de 
Gen\`eve,\\ 24 quai Ernest 
Ansermet, CH--1211 Gen\`eve 4, Switzerland\\
 \email{marc.vonlanthen@unige.ch}
 }
 \author{Ruth Durrer
\\D\'epartement de Physique Th\'eorique and Center for Astroparticle Physics,\\
Universit\'e de 
Gen\`eve,\\ 24 quai Ernest 
Ansermet, CH--1211 Gen\`eve 4, Switzerland\\
 \email{ruth.durrer@unige.ch}
 }
\received{\today}

\abstract{
We study the distance--redshift relation in a universe filled with 'walls' of pressure-less dust
separated by under dense regions. We show that as long as the density contrast of the walls is small, or the diameter of the under dense regions is much smaller than the Hubble scale, the distance--redshift relation remains close to what is obtained in a Friedmann
universe. However, when arbitrary density contrasts are allowed, every prescribed  
distance--redshift relation can be reproduced with such models.
}
\begin{document}
\maketitle

%\author{ {\color{blue}Enea} Di Dio,  {\color{green}Marc} Vonlanthen and {\color{red}Ruth} Durrer }
%\affiliation{D\'epartement de Physique Th\'eorique and Center for Astroparticle Physics, 
%Universit\'e de 
%Gen\`eve\\ 24 quai Ernest 
%Ansermet, CH--1211 Gen\`eve 4, Switzerland}

%\preprint{ }

%\date{\today}

\section{Introduction}\label{s:intro}

Since more than a decade, cosmology research is facing the dark energy problem:
the present Universe seems to be in an accelerating phase. This conclusion was
first drawn from measurements of the distance--redshift relation from type Ia Supernovae 
(SNIa)~\cite{SNorig,SNnew} and is confirmed by many other datasets, from
the cosmic microwave background~\cite{cmb} to baryon acoustic oscillations and other
aspects of large scale structure.
Until very recently the measurements inferring the existence of dark energy rely mainly on the
distance--redshift relation which is valid in a Friedmann Universe~\cite{roysoc}. New
{\em independent} measurements of, e.g. the expansion rate $H(z)$ are now being performed
see e.g.~\cite{arXiv:1108.2637}. Hence this situation is changing, so that we shall soon  know both,
$d_A(z)$ and $H(z)$ with good accuracy. The general opinion is
that fluctuations on large scales are small so that they can be treated 
with linear perturbation theory and linear perturbations average out in the mean over
many directions and large scales, and therefore fluctuations are not relevant for the determination of 
quantities like $d_A(z)$ and $H(z)$. This expectation has been confirmed by perturbative calculations.
Within  linear perturbation theory, the fluctuations of the distance--redshift relation 
for redshift $z>0.2$ is on the level of a few percent~\cite{CRA}.

However, perturbations on smaller scales can become very large, density fluctuations
e.g. in galaxies are $\de\rho/\rho \sim \rho_{\rm gal}/\rho_m \sim 10^8$. Since the 
relation between metric perturbations, or more precisely the Christoffel symbols, and 
density fluctuations is non-linear, it is not evident that large amplitude, non-linear, small scale 
density fluctuations cannot add up to affect the distance--redshift relation on large scales.

To study the real problem one would need to analyse light rays passing through a realistic
Universe with high density fluctuations. So far, this has been done only within Newtonian N-body 
simulations, see e.g.~\cite{Nbody}. However, it is well known that Newtonian gravity misses 
the terms which are relevant for the back reaction problem~\cite{newton}, hence a full, 
non-linear relativistic treatment is needed. Since this is very difficult, so far mainly toy models
which mimic reality to a certain extent have been studied. 

The present work inscribes in
this framework. Instead of considering spherically symmetric solutions of general relativity (GR),
the so called Lema\^\i tre~\cite{L}-Tolman~\cite{T}-Bondi (LTB) models, for recent reviews
see~\cite{LTB}, we study
a Universe containing high density walls. We shall consider infinitely extended parallel walls.
The considered model is a sub-case of the Szekeres solution~\cite{szekeres}. Light propagation in general Szekeres model has been studied recently~\cite{Kras2,marco}.
This is of course a gross over-simplification, but we know that galaxies tend to be aligned in
filaments and photons coming to us from a far away supernova, might experience a geometry
similar to the one of such a symmetric wall universe. The weakest point of our toy model is that
all the walls are parallel while we expect a typical photon to traverse filaments which are 
aligned in different directions. We shall  take this into account to some extent by studying 
photons coming in from  different directions with respect to the  walls.

Such walls have been studied in the past~\cite{collins}, but only perturbatively. Since we know
that the effects are small within linear perturbation theory, we cannot trust higher order
perturbation theory if it predicts large deviations from the Friedmann distance-redshift relation.
For this reasons we analyse exact, fully relativistic  wall-universes in this work.

In the next section we present the wall metric and the Einstein equations. We also study the 
conditions on the parameters which have to be satisfied so that no singularity apart from
the Big Bang is present in the backward light cone of the observer. In section~\ref{s:res} 
we present the results for the distance-redshift relation for 'realistic' walls and for a wall 
universe which mimics the observed relation. In section~\ref{s:con} we conclude.

\section{Wall Universes}\label{s:eq}
In this section we study universes containing only pressure-less matter (dust) and which are 
symmetric under translations and rotations in a plane which we call the $y$-plane.
They have the same number of symmetries as LTB models and can be solved
analytically, see~\cite{zakharov}.
The metric is of the form
\be\label{e:metric}
ds^2 = -dt^2 + a^2(t,x)dx^2 + b^2(t,x)(dy_1^2+dy_2^2)\, .
\ee
Note that  the only difference to the LTB geometry is that our symmetrical 2d manifolds are
planes, $dy_1^2+dy_2^2 = dr^2 + r^2d\phi^2$ while those of LTB are spheres, $d\Om^2= d\theta^2 +\sin^2\theta d\phi^2$. We denote the spatial coordinates by $\bx = (x,y_1,y_2)$
in order to reserve the letter $z$ for the redshift. In the following a prime denotes a derivative 
w.r.t. $x$ while  a dot denotes 
 derivative w.r.t. $t$. The Einstein equations for this geometry and for pure dust matter 
 yield~\cite{szekeres,zakharov,kras}
\bea
\dd_t\left(\frac{b'}{a}\right) \equiv \dd_tE &=& 0 \label{e:0i}\, , \\
\dot b^2 - \left(\frac{b'}{a}\right)^2 &=& 2   \frac{M(x)}{b} \, , \label{e:00}\\
M' = 4\pi G\rho b^2 b' &=&  4\pi G\rho b^2aE(x)  \label{e:M'}\,.
\eea
In Eq.~(\ref{e:0i}) we have introduced the time-independent function 
\be\label{e:E}
E(x) = b'/a
\ee
and Eq.~(\ref{e:00}) defines $M(x)$ which is also 
time-independent.  In LTB models $M/G$ can be interpreted as mass 
density (Note that in the LTB case a term $b/(2G)$ has to be added to $M$ 
which is a consequence of the 
curvature of the 2-sphere. For more details see~\cite{kras}.), and $(M'/G)r^2dr$ is the mass in a shell 
of thickness $dr$.  However as the mass in an infinite plane is not well defined, this 
interpretation is not meaningful in the planar case. In our case it is therefore not unreasonable that $M$ may become negative even though $a$, $b$ and $\rho$ are supposed to be 
positive at all times.

From the matter conservation equation we also obtain $\dd_t(\rho b^2 a)=0$, which, on the other
hand, is a consequence of Eq.~(\ref{e:M'}).

\subsection{The solutions}
Eq.~(\ref{e:00}) can we rewritten as
\be\label{e:b}
\dot b^2 = \frac{2M(x)}{b} +E(x)^2 \, ,
\ee
with parametric solutions~\cite{szekeres,zakharov}
\bea
 %&&  \mbox{ for } E\neq 0~ : \nonumber\\
 \mbox{ for } E\neq 0~ : \qquad b &=& \frac{M}{E^2}(\cosh\eta -1)  =  \frac{2M}{E^2}\sinh^2(\eta/2) \label{e:En0Mp} \, , \\
t &=& \frac{M}{E^3}(\sinh\eta -\eta)+t_B(x), ~  \mbox{ for } M>0 \, ;  \label{e:En0Mpt}\\
b &=& -\frac{M}{E^2}(\cosh\eta +1)   =  -\frac{2M}{E^2}\left(\sinh^2(\eta/2) +1\right) \, ,
\label{e:En0Mn}\\
t&=& -\frac{M}{E^3}(\sinh\eta +\eta)+t_B(x) , ~ \mbox{ for } M<0 \, ;\\
b &=& |E|(t-t_B(x)) \quad \mbox{ for }  M=0 \, ; \label{e:M=0}\\  && \nonumber \\
 %&&  \mbox{ for } E= 0~ : \nonumber\\
\mbox{ for } E= 0~ :\qquad   b &=& \left(\frac{3}{2}\sqrt{2M}(t-t_B(x))\right)^{2/3}, \mbox{ for }  M>0 \,, \label{e:E=0}\\
b &=& b_0 = \mbox{ const. },  \mbox{ for }  M=0 \, . \label{e:EM=0} 
\eea
Note that for  $E=0$ Eq.~(\ref{e:b}) implies that $M\ge 0$. This equation also implies 
$$ b \ge -\frac{2M}{E^2} $$at all times, in all cases.
 
The function  $t_B(x)$ is arbitrary; it is called the 'bang time'. For $M\ge 0$, at
$t=t_B$, i.e $\eta=0$, we have $b=0$ which represents the Big Bang singularity. 
Positions with $M<0$ have no Big Bang singularity but a 'bounce' at $t=t_B$. We 
shall simplify below to the case
$t_B\equiv 0$, i.e., uniform bang time. Note that we have chosen expanding solutions.
From these we can  obtain the collapsing solutions simply by changing the sign of $t$.
Since in the Einstein equations only $\dot b^2$ appears they are invariant under $t\ra -t$. 

Of course the $\{t = $const.$\}$ hypersurfaces are not parallel to the $\{\eta = $const.$\}$ 
hypersurfaces, but their position depends on $x$. For fixed position $x$,  
Eqs.~(\ref{e:En0Mp},\ref{e:En0Mpt}) and (\ref{e:E=0}). correspond to Friedmann 
solutions with curvature $K= -E^2\le 0$ and $M=4\pi G\rho b^3/3$.  Note that unlike in the 
Friedmann case, wall solutions with $M<0$ need not be unphysical.

The parametric representation with $\eta$ is chosen in order to express the solutions
in terms of elementary functions, but it is of course not necessary.
For example, for $M>0$, setting
$$\tau(t,x) = -E^2 \left(\frac{t}{6M}\right)^{2/3} \  \mbox{ and}$$
$$
 S(\tau) = (-3\tau)^{-1}\sinh^2\left(\frac{1}{2}\left[\sinh-\rm{id}\right]^{-1}\!\!\left(6\left( -\tau\right)^{3/2}\right)\right)
$$
we obtain
$$ b(t,x) = -\frac{M}{E^2}6\tau S(\tau) \, . $$
Note that in the definition of $S$, $\left[\sinh-\rm{id}\right]^{-1}$ denotes the inverse of the
function in brackets, and $\rm{id}$ is the identity function, $\rm{id}(x)=x$.
One can check that $S$ solves the differential equation~\cite{yoo} 
\begin{equation}
\frac{4}{3} \left( S+\tau S' \right)^2 +3\tau - \frac{1}{S} =0 , \label{8}
\end{equation}
with initial condition $S(0)=\left( \frac{3}{4} \right)^{1/3}$. Note that this is the only regular solution,
i.e solution with $S'(0) \neq \infty$.
This expression will be useful in Section~\ref{s:mimi}.

The function $a(x,\eta)$ can be obtained from Eq.~(\ref{e:E}). For example for $M>0$
 we find
\begin{eqnarray}
& a =& E^{-1}\left(\frac{\dd b}{\dd x}\right)_t  \nonumber \\ && \nonumber \\ 
 \mbox{ for } E\neq 0~ : \quad &
a =& 
\frac{2}{E}\left( \frac{M}{E^2} \right)' \sinh^2\left( \frac{\eta}{2}\right)\!\! - \coth\left( \frac{\eta}{2} \right) \Big[t'_B  +  \left( \frac{M}{E^3} \right)' \!\!\left( \sinh\eta\! -\! \eta\right)\! \Big] \label{e:a}
  ,  \label{e:aE}\\
   \mbox{ for } E = 0~ : \quad &
a =&  \frac{(t-t_B)^{2/3}}{M^{1/3}6^{1/3}}\!\left[M^{-1/3}\frac{M'}{E} + 
    \frac{9(t-t_B)^{2/3}E' }{5\times6^{1/3}}
 \right]\!  .  \label{e:a0}
 \end{eqnarray}
(The suffix $t$ in $\dd b/\dd x$ indicates that we have to interpret  $b$ as functions of $(t,x)$, 
not $(x,\eta)$, in this derivative.) Even if $E=0$, Eq.~(\ref{e:M'}) implies that 
$0<M'/E<\infty$, so that the r.h.s. of Eq.~(\ref{e:a0}) is well defined. Below, we shall choose
the $x$-coordinate such that $M'/E=$constant.
 
Note that  $M(x)$ and $E(x)$ can pass through zero so that in general different solutions from
above have to be glued together at the boundary of their validity. We have checked that 
this gluing process can be performed in a smooth way and does not induce 
singularities in the scale factor $b$. However, for $M \rightarrow 0$ the scale factor $a \rightarrow \infty$. 
Nevertheless, we believe this to be a coordinate singularity, since, as we have checked,  
both, the Kretschmann scalar, $K\equiv R_{\al\beta\mu\nu} R^{\al\beta\mu\nu}$  and the 
scalar curvature remain finite for $ M \rightarrow 0$. In our examples below we shall have $M>0$ throughout and therefore we do not encounter this problem. However, when computing $a$ 
from Eq.~(\ref{e:E}), one has to be careful to use the result (\ref{e:aE}) and take 
the limit $E\ra 0$ for fixed $t$, hence also $\eta\ra 0$.
One cannot use (\ref{e:E=0}) and (\ref{e:E}), since for $E=0$ we have $M'=0$ so that
Eq.~(\ref{e:E}) is identically satisfied and cannot be used to obtain $a(t,x)$.

\subsection{Singularities}
Singularities can occur when $a$, $b$ or $\rho$ become either infinite or zero.
To have no singularities (apart from the Big Bang) which occurs at
$t=t_B$, hence $b=0$, in the past light cone of every 
possible observer we might be interested in, we must demand that all singularities lie in
the future. In more precise models, when one specifies the observer location,
one can relax this condition to the one that no singularity lies within the background lightcone
of the specific observer. 

In general, the question of singularities depends on the choice of the functions $M(x)$ 
and $E(x)$. From our solutions it is clear that $b$ behaves monotonically as a function 
of time for fixed $x$. This is to be 
expected since no clustering goes on in the directions $y_1$ and $y_2$ described by this scale 
factor. Since we are interested in an expanding $b$, a singularity is present when the
the scale factor $a$ of the $x$-direction tends to zero. From Eq.~(\ref{e:a}) we infer 
that  for $t_B\equiv 0$, $a=0$ implies
$$  \frac{2}{E}\frac{(M/E^2)'}{(M/E^3)'} = \frac{\cosh(\eta/2)}{\sinh^3(\eta/2)}
\left(\sinh\eta -\eta\right) \ge 4/3\,.$$
It is easy to verify that the right hand side is an even positive function with minimum $4/3$
at $\eta=0$. Hence there is a singularity at some finite value of $\eta$ if the l.h.s. ever
becomes larger than $ 4/3$ or, equivalently, if
$$ \frac{E'}{E}\frac{M/E^3}{(M/E^3)'} =\frac{\cosh(\eta/2)}{2\sinh^3(\eta/2)}
\left(\sinh\eta -\eta\right) -1 > -1/3$$
for some value of $x$.

We now consider a simple ansatz motivated by the perturbative analysis presented
in Ref.~\cite{zakharov}. We choose
\begin{equation}\label{e:Mh} 
M(x)= \frac{2}{9t_0^2}\left( 1 + \epsilon h( x ) \right)
\end{equation}
and 
\be
4\pi G\rho b^2 a = \frac{M'}{E} = \frac{2}{3}t_0^{-2} = \mbox{const.}
\ee
so that
\be\label{e:Eh}
E = \epsilon \frac{h'}{3}\,.
 \ee
 
 In full generality $M'/E =f(x)$ could be an arbitrary positive function of $x$. But we can always
 make a coordinate transformation to $\tilde{x}(x)$ determined by
 $$ \frac{dx}{d\tilde x} =\frac{1}{6\pi G\rho b^2 at_0^2} , $$
 so that with respect to the new 
 coordinate $M'/E =$constant. Hence we just fix the coordinate $x$ (up to a constant shift) 
 by this choice. In addition, we
 have chosen uniform bang time, $t_B(x)\equiv 0$. This is a true restriction. With this
we have reduced the three free functions of $x$ to one, $h(x)$ which 
defines the density profile. Furthermore, we have introduced the parameter $\ep$
such that for $\ep=0$ we reproduce the matter dominated Friedmann solution. We may
also require $|h(x)|\le 1$ so that $\ep$ indicates the amplitude of the perturbations.
We do this in one of the examples below.

The above requirement for a singularity at some time $t\neq 0$ now reduces to
$ MM'' <M'^2/3 $. (Strictly our derivation applies only for $M'\neq 0$. For 
$M'\propto E =0$, one sees directly from Eq.~(\ref{e:a0}) that
$M'' \propto E'<0$ is the necessary and sufficient condition for $a=0$ at some time $t>t_B$.)
We have found that most interesting mass profiles satisfy this
condition for some values of $x$ and therefore have singularities at some time 
in some places. This is not surprising but actually expected from gravitational collapse. 
However, when over densities become very high and we approach the collapse, pressure 
forces and heating become important and our simple pressure-less dust model for matter no longer holds. In order to be able
to stay within the present framework, we therefore demand that such singularities be in the future
and not in the past for the density profiles under consideration.

Let us consider as a first example 
$$ h(x) = \cos(kx) \,. $$
Then the condition for the existence of a singularity (at $t\neq 0$) becomes
$$ -\left(\ep\cos(kx) +\ep^2\cos^2(kx)\right) <(\ep^2/3)\sin^2(kx) \, ,$$
which is always satisfied for some values of $x$, irrespective of $k$ and $\ep$. 
A similar behavior is expected 
whenever $h$ is not a convex function, but a function representing several under- 
and over-densities cannot be convex.

However, this is not so important for our considerations. As we have said, the requirement of singularities to be absent is mainly a technical one and it is actually sufficient not to have a singularity in the past.

Using the above expression for $a$ (for $M>0$) and the ansatz (\ref{e:Mh},\ref{e:Eh}) 
for $M$ and $E$, we find that $a=0$ is equivalent to
\bea 
\frac{(1+\ep h)h''}{\ep h^{\prime 2}-3(1+\ep h)h''}  = 
-\frac{1}{3}\frac{1}{1-\frac{\ep h'^2}{3(1+\ep h)h''}}  = \frac{1}{2}\frac{\cosh(\eta/2)}{\sinh^3(\eta/2)}\left(\sinh\eta -\eta\right)-1 >-1/3  \,. 
\label{e:hcond}
 \eea
Interestingly, in extremal positions of $h$, with $h'=0$, the l.h.s. of the above expression
is exactly $-1/3$. This comes from the fact that for this case $\eta=0 \ \forall \  t$ and
we have to replace the condition that there is no singularity before some given time $t_0$
by $a(t)>0$ for $t<t_0$ using expression (\ref{e:a0}) for $a(t)$.  If $h'''=0$ when $h'=0$ (as in our example)
one can show that in the
positions where $h$ has a maximum, hence $h'=0$ and $h''<0$, $1+\ep h>0$,  
singularities occur first.
 Furthermore, when $1+\ep h>0$ and $h''<0$,  the denominator of the l.h.s. of Eq.~(\ref{e:hcond}) is larger than $1$ and hence the l.h.s. becomes larger than $-1/3$. Therefore, there exists a finite value $\eta_s(x)$
 where Eq.~(\ref{e:hcond}) is satisfied and $a(x,\eta_s(x))=0$. If, on the contrary,  $1+\ep h>0$ and $h''>0$ 
 the l.h.s. of  Eq.~(\ref{e:hcond}) is smaller than $-1/3$. For positions in the vicinity of an extremum
this implies that if the extremum is a minimum of $h$, the position $x$ does not encounter a singularity in the future
while positions close to maxima do.

\begin{figure}[h!]
\centering  ~  
%\vspace{1cm}
\includegraphics[clip=true, width=12cm]{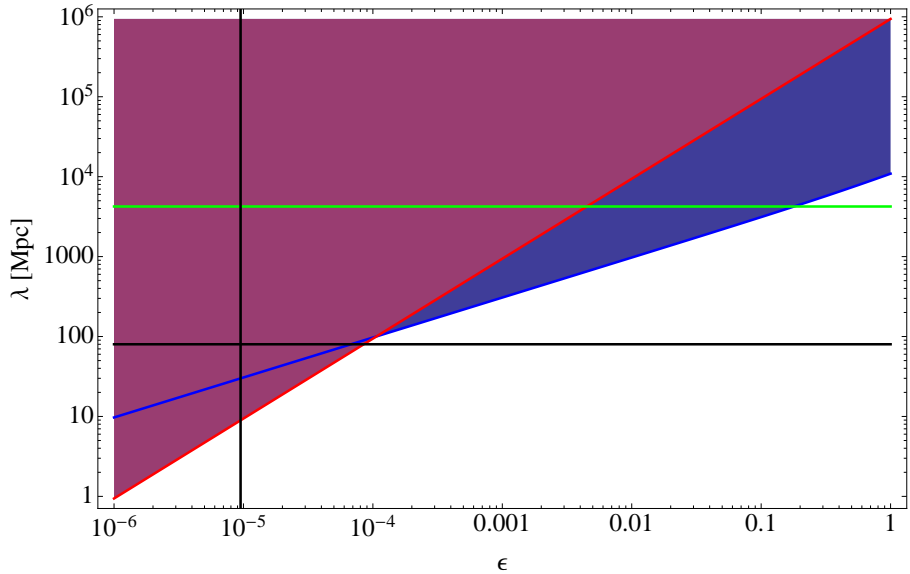}
\caption{\label{fig1}
The region above the red line has singularities in the future only. While the blue line describes the condition to use the  limiting solution for $E=0$, (\ref{e:a0}). This can be used when $tE^3/M\ll 1 $, where with "$\ll$" we mean at least two orders of magnitude smaller. The two black lines describe the physical parameters $\epsilon = 9.5 \times 10^{-6}$ and $\lambda = 80$ Mpc. The green line is the Hubble scale $H_0^{-1}$. With physical parameters we mean an amplitude as determined by WMAP~\cite{WMAP} observations and a wavelength 
agrees with the size  of  the largest observed voids~\cite{largevoid} which is about $40$-$90$ Mpc. More precisely we find $\epsilon$ requiring that at early time there is only a single density fluctuation in each Hubble distance. This leads, at first order, to $\delta= 8 \pi^2 \epsilon /15$, and the matter density fluctuation at early times, 
$\delta\cong 5 \times 10^{-5}$ can be inferred from WMAP observations. For more details see~\cite{collins}.
}
\end{figure}

Let us study in more detail the request that the second singularity (not the big bang one) lies
in the future, $t>t_0$. Using the expression~(\ref{e:En0Mpt}) for
$t$, we can rewrite the condition $a(x,\eta_s)=0$ as 
$$\frac{(1+\ep h)h''}{\ep h^{\prime 2}-3(1+\ep h)h''}  =  \frac{9}{4}\frac{\cosh(\eta_s/2)}{\sinh^3(\eta_s/2)}\frac{t_0^2t(x,\eta_s)\ep^3 h'^3}{(1+\ep h)} -1 \,.$$
The condition $t(x,\eta_s)>t_0$, for  $h'<0$ which we shall consider
hence $\eta_s<0$ for $t(x,\eta_s)>0$,  then becomes
$$ \frac{(1+\ep h)}{\ep^3 h'^3}\left[ 1 + \frac{(1+\ep h)h''}{\ep h^{\prime 2}-3(1+\ep h)h''} \right]
\frac{4}{9t_0^3} < \frac{\cosh(\eta_s/2)}{\sinh^3(\eta_s/2)} \,.$$ 
This equation for $\eta_s(x)$ can only be solved numerically. However, often we realize that
the l.h.s. is smallest at small $\left| h' \right| $ i.e. for small values of $\left| E(x)\right| $. Hence singularities
will develop first in positions with small $\left|h' \right|$.  This requires also small $ \left| \eta_s \right| $ so that we
may develop the scale
factor $a$ and $t$ in $\eta_s$. The above inequality then leads to power law relations and 
inserting the above expression for $E=(3/2)M't_0^2$ yields the constraint
\begin{eqnarray} \label{7}
 &&1+ \frac{(3 t_0^2)^{7/3}3^{1/3}2^{2/3}}{80} \left( 6 M'' M^{1/3}  - \frac{M'^2 }{M^{2/3}} \right)  >0 \, , \nonumber\\
 && 1- \frac{1}{20} (t_0k)^{2/3} \Big(6\epsilon \cos( k x ) \left( 1 + \epsilon \cos(k x) \right)^{1/3} 
   +\frac{\epsilon^2 \left( \sin\left(k x \right) \right)^2}{\left( 1 + \epsilon \cos \left( k x \right) \right)^{2/3} } \Big) > 0 .
\end{eqnarray}
The first inequality is general while for the second inequality we have chosen $h=\cos(kx)$.
In Fig.~\ref{fig1} we plot the constraint for this case together with the condition to use the limiting solution for $E=0$, (\ref{e:a0}),
(which is not necessary for our analysis) in the $\ep$--$\la$ plane, where
$\la$ denotes the wavelength of the perturbation $\la=2\pi/k$.
 
\section{The distance redshift relation in a wall universe}\label{s:res}
\subsection{Generalities}
\subsubsection{Redshift} 
We now consider a photon emitted from a source at some position and 
time $(t_s, \bx_s)$ 
arriving in our telescope at position and time $(t_0, \bx_0)$. We denote the matter 4-velocity
 field, hence the 4-velocity of source and observer by $u(t, \bx)$ and the photon 4-velocity 
 by $n$. The redshift of the source, $z$ is then given by
 \be
 1+ z = \frac{g(n,u)|_s}{g(n,u)|_0} \,.
 \ee
We consider a co-moving source and observer, hence $u=\dd_t$ and normalize the affine
parameter of the photon, $s$, such that $n^0(s_0)=1$. The redshift then reduces to 
\be
1+ z = n^0|_s
\ee 
for our geometry with $g_{00}=-1$ and $g_{0i}=0$. From the geodesic equation for the photon
we infer that its momenta in $y_1$- and $y_2$-direction are simply redshifted so that
\bea
 J_1 \equiv b^2n^1 = b^2 \frac{dy_1}{ds} = \mbox{const. ~~~  and} \qquad 
 J_2 \equiv b^2n^2 = b^2 \frac{dy_2}{ds} = \mbox{const.}
\eea 
hence
\be\label{e:k1}
 (n^x)^2  = \left( \frac{n^0}{a} \right)^2 - \frac{1}{a^2b^2} \left( J_1^2 + J_2^2 \right). 
\ee
From the geodesic equation for $n^0$ we can now derive the evolution of the redshift:
\be\label{e:zs}
\frac{dz}{ds}= -\frac{dn^0}{ds} = (1+z)^2\frac{\dot b'}{b'} + \frac{J_1^2+J_2^2}{b^2}
\left( \frac{\dot b}{b}- \frac{\dot b'}{b'} \right) \,.
\ee
Here we have used $a=b'/E$ to eliminate the scale factor $a$. Note also that the prime and the dot in the above equation denote partial derivatives while $d/ds$ is a total derivative along the 
path of the photon.

\subsubsection{Distance}
The evolution of the distance to the source is given by the Sachs focussing equation~\cite{sachs},
\be\label{e:foc}
\frac{d^2D}{ds^2} = -\left(|\si|^2 + \RR\right)D \,.
\ee
$D$ is the angular diameter distance to the source, $\si$ is the complex scalar shear of the light
bundle which we define below and
\be
\RR = \frac{1}{2}R_{\mu\nu}n^\mu n^\nu = 4\pi GT_{\mu\nu}n^\mu n^\nu = 4\pi G(1+z)^2
    (\rho + \bar P) \,. 
\ee
Here $\bar P$ is the pressure in the direction of the photon. The important point
is that this quantity is non-negative for any energy momentum tensor which satisfies the dominant
energy condition $\rho\ge \bar P$ in all directions, hence also for a cosmological constant 
where we have $\RR\equiv 0$. In terms of the affine parameter of the photon, the growth of the
angular diameter distance to the source is not accelerated. If the dominant energy condition is satisfied $D(s)$ is always a concave function. Furthermore, clustering which leads to the
production of non-vanishing shear is only increasing the deceleration of $D$ as function
of the affine parameter $s$. But of course we do not measure this
function but $D(z)$ which can behave very differently.  

The complex shear of the light ray bundle is defined as follows (appendix 7.7.3. in~\cite{ns}): We consider
two spatial orthonormal vectors $e_1$ and $e_2$ which are normal to both, $u$ and $n$ 
at the observer and are parallel transported along $n$, such that $\nabla_n e_a =0$ for $a=1,2$.
The vectors $e_1,e_2$ are a basis of the so called 'screen'. Note that we do not require
that $u$ be parallel transported along $n$, hence $e_1,e_2$ are in general not normal 
to $u$ elsewhere than at the observer, where we have given their initial conditions.
The complex shear is defined by
\be\label{e:shear}
\si = \frac{1}{2}g(\ep,\nabla_\ep n)\, , \qquad \ep \equiv e_1 + i e_2.
\ee
In order to compute the shear we must know $n$ not only along the photon geodesic itself
but we must determine its derivatives in directions normal to $n$. We shall directly use the
transport equations~\cite{ns}. For a vorticity free ray bundle (which is the case here) with
expansion rate $\theta \equiv \frac{1}{2}n^\mu_{;\mu}$ these are
\begin{eqnarray}
\dot \theta  + \theta^2 + \sigma_1^2 + \sigma_2^2 &=& -\RR, \label{e:170}\\
\dot \sigma_1 + 2 \theta \sigma_1 &=& - \text{Re} \left( \FF \right), \label{e:171}\\
\dot \sigma_2 + 2 \theta \sigma_2&=&  \text{Im} \left( \FF\right), \label{e:172}
\end{eqnarray}
where $\sigma_1= \text{Re} \left( \sigma \right) $, $\sigma_2= \text{Im} \left( \sigma \right)$, and $\FF= \frac{1}{2} R_{\alpha \mu \beta \nu} \bar \epsilon^\alpha \bar \epsilon ^\beta n^\mu n^\nu$.
To determine the shear $\sigma$ we need to know the initial conditions for the differential equations (\ref{e:170}) to (\ref{e:172}). It is possible to determine the behavior of the shear and the expansion of the light near the vertex~\cite{SPE}. Choosing the affine parameter
of the photon to vanish at the observer position, $s_0=0$, these are
\begin{eqnarray}
\sigma(s)&=&-\frac{s}{3} { \bar \FF}_0+ O \left(s^2\right),\\
\theta(s) &=&\frac{1}{s} \left( 1 - \frac{1}{3} \RR_0 s^2 \right) + O \left( s^3 \right).
\end{eqnarray}
$\FF_0$ and $\RR_0$ are the values of $\FF$ and $\RR$ at the observer position.
The light bundle expansion $\theta$ diverges at the observer position, but we can consider an initial condition not exactly at the observer. This choice can affect the numerical precision.
After determining $\RR$, $\ep$ and $\FF$ for a given geometry and photon direction, 
we can solve the system (\ref{e:170}) to (\ref{e:172}) together with the Sachs focusing 
equation (\ref{e:foc}) numerically.

%%%%%%%%%%%%%%%%%%%%
\subsection{'Realistic' walls}
We want to investigate whether the system of equations derived above for $z(s)$ and $D(s)$
can lead to a distance-redshift relation close to the one observed. For wall universes we consider,
\be\label{e:Rab}
\RR = 4\pi G \rho (1+z)^2= \frac{2(1+z)^2}{3t_0^2b^2a} \;.
\ee
For a chosen density contrast $h(x)$ we can determine $b(t,x)$ and $a(t,x)$ and solve
the photon geodesic Eq.~(\ref{e:zs}) for a given angle $\theta_0$ of the observed photon w.r.t. 
the $y$-plane,
\be\label{e:theta} 
\cos\theta_0 = \frac{\sqrt{J_1^2+J_2^2}}{b(x_0,t_0)n^0(0)} \, .
\ee
We again set the initial value or the affine parameter to $0$, hence $x_0=x(0)$ etc.

\begin{figure}[h!]
\centering  ~  
%\vspace{0.5cm}\\ 
\includegraphics[clip=true, width=12cm]{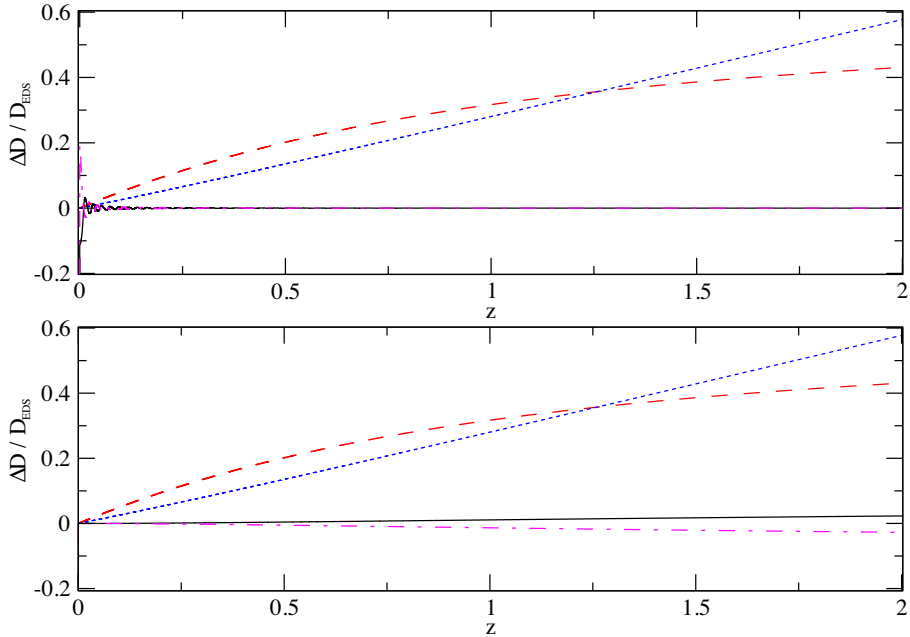}
\caption{\label{f:cos}We show the relative luminosity distance redshift relation $\frac{\Delta D(z) }{D_{EdS}(z)} = \frac{ D(z)-D_{EdS(z)} }{ D_{EDS}(z)}$,
for different models with luminosity distance $D(z)$. The blue dotted curve is for a Milne Universe, the red dashed curve is for $\La$CDM universe with
$\Omega_{\Lambda}=0.7$ and $\Omega_{M}=0.3$. The remaining two lines are our wall universe. The 
black solid line is in an under density while the purple dot-dashed line is in an over density. In the top panel,
we consider light propagating
in the $x$-direction only.  The bottom 
panel is the same but for light propagating in the $y$-direction. The parameters for the wall model are the physical ones, $\epsilon = 9.5 \times
10^{-6}$ and $\lambda = 80$ Mpc.
}
\end{figure}

We have investigated two choices for $M(x)$. The first is simply $M(x)=\frac{2}{9t_{0}^{2}}\left(1+\ep\cos(kx)\right)$
which we have already discussed before. The results for this case are shown in Fig.~\ref{f:cos}.
 
The result is quite striking: The deviation from the Einstein-de Sitter distance-redshift relation is very small. 
On the level of a few percent in the most extreme case. Much smaller than the deviation for an 
open (Milne) Universe or even for $\La$CDM.
Hence voids and walls with the chosen parameters cannot simulate the observed distance redshift relation.
We have also studied different values of the parameters $(\ep,k)$, but all cases which are such that
there is no singularity before $t_0$ lead to small deviation from Einstein-de Sitter. Only for wavelengths of approximately Hubble scale, $k\sim H_0$, where we can choose $\ep \sim 10^{-3}$ do the deviations become relatively
large. But the density profile chosen here does not at
all lead to a relation that resembles the observations.

\begin{figure}[h!]
\centering  ~  
%\vspace{1cm}\\ 
\includegraphics[clip=true, width=12cm]{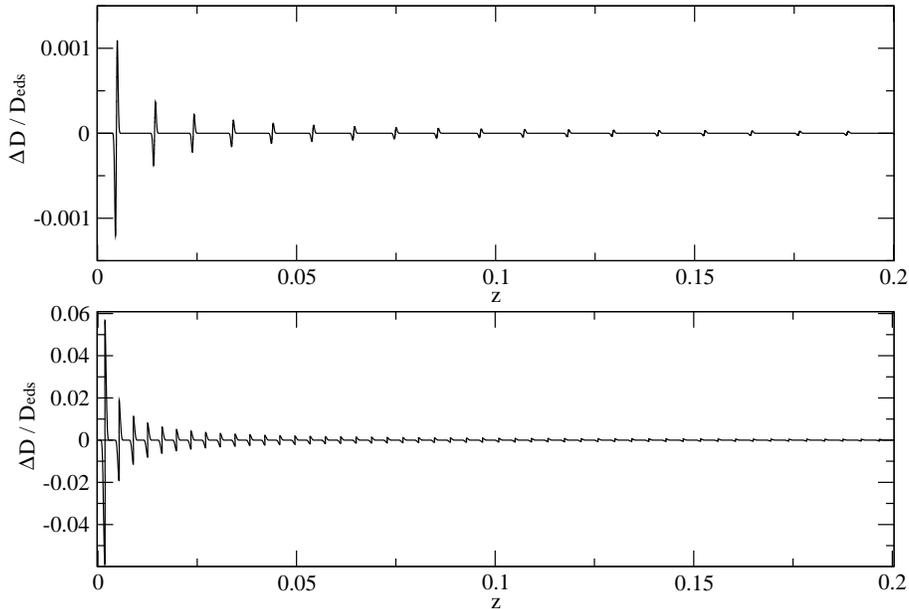}
\caption{\label{f:exp} We show the relative difference between the distances in 'realistic' wall models and in EdS universe for photons propagating in $x$-direction. The top panel is obtained with $\epsilon =10^{-9}$, $\lambda= 40 $ Mpc and $\sigma=1$ Mpc, while the bottom one with $\epsilon=5 \times 10^{-8}$, $\lambda= 15 $ Mpc and $\sigma=1 $ Mpc. In both cases the observer is at the center of the void. We have checked that the order of magnitude does not change for an observer in an over density. In the second case, we see that we obtain an effect of the same order of magnitude as
the swiss cheese universe discussed in~\cite{Marra}. }
\end{figure}

As a second profile we consider thin, highly concentrated over-dense walls with an
exponential profile:
\be\label{e:exp}
h \left(x \right) = \frac{\lambda}{\sqrt{2 \pi \sigma^2}}\sum_i \exp\left(\frac{-(x-x_i)^2}{2\si^2}\right)  -1, 
\ee
where $\la = x_{i+1}-x_i$. In the limit $\sigma \ll \lambda$ the mean of $h\left( x \right)$ vanishes and $\text{min}_{x} h \left( x \right)  =-1$.
Again, we choose $\ep$ such that there is no singularity before $t_0$. The results for this profile are shown in Fig.~\ref{f:exp}.

We have obtained the following result in these two examples (and other profiles 
which we do not present here explicitly): The modification of $D(z)$ never goes beyond
the case of the open universe. We do not obtain acceleration by a series of dense walls.
Even though we present here only two simple profiles, we think the conclusion is
valid beyond these cases: if a photon passes through many {\em compensated} under-
and over-densities in the integrated distance $D(z)$ the effect is minute as long as the time the photon spends inside a wall is much smaller than the time scale at which the gravitational potential of the wall evolves. A perturbative (first order)
calculation gives a flavour of this effect. Indeed, at first order in the perturbed direction,
the difference between $D(z)$ in our models and $D_{EdS}(z)$ of a matter dominated 
universe can be written as
\begin{eqnarray} \label{linearregime}
D_L \left( z_e \right)-D_L^{EdS} &=&\left( 1 +  z_e \right) \left( \eta_o -\eta_e \right) \left( \frac{\epsilon}{3} \left( h \left( \eta_o \right) + h \left( \eta_e \right) \right)  \right)   - \left( 1 +  z_e \right)  \int_{\eta_e}^{\eta_o}    \frac{2\epsilon}{3} h \left( \eta \right) d\eta\nonumber \\
&&  \hspace{-1.3cm}+ \left( 1 +  z_e \right)   \int_{\eta_e}^{\eta_o} d\eta \int_{\eta_e}^\eta d\eta' \frac{\epsilon}{15} h'' \left( \eta' \right) \eta'      -   \frac{  1 +  z_e }{\mathcal{H}_e}   \int_{\eta_e}^{\eta_o} d\eta \ \frac{\epsilon}{15} h'' \left( \eta \right) \eta , 
\end{eqnarray}
where the subscripts $e$ and $o$ respectively mean that the conformal time is evaluated 
at the source (emission) or at the observer and expresses the perturbation of the energy density in under 
and over densities (see Appendix C for
a derivation of the linearized result). From this expression, valid in 
the linear regime only, and for a periodic perturbation, it becomes clear that 
the deviation of $D_{L}(z)$ with respect to $D_{L}^{EdS}$ depends on the amplitude $\epsilon$ of the 
perturbation and on the values of the conformal time at the source and at the observer. 
In the case of periodic perturbations,
 the contributions from photon path are mostly cancelled in the integral terms. 
Of course in the full non-linear calculation there is no simple relation between the matter over density $h$ and the
gravitational potential. In this case in principle the full non-linear Einstein equation have to be solved and
Eqs.~(\ref{e:zs}) and~(\ref{e:foc}) govern $D_L(z)$.

Surprisingly, however,
our non-linear simulations show that this result holds also to some extent
in the non-linear regime. Note that, even though our value of $\ep$ is small, the over densities in 
the walls are large at late times, such that they develop singularities soon after today and we are deeply in the 
non-linear regime. While we do not have a proof that our conclusion holds in all cases, we have tested this also with 
other periodic wall profiles.

In Fig.~\ref{f:exp,Hubble} we show the deviations of the expansion rates with respect to the Hubble expansion in EdS universe. We note that the deviations in the unperturbed directions are small. However, in the perturbed direction these deviations can be large locally inside a wall, and they would be measurable by direct, local measurements of $H(z)$.
However, they compensate when averaged over a wall thickness and do not show up in integrated quantities like $D(z)$.

\begin{figure}[h!]
\centering  ~  
%\vspace{1cm}\\ 
\hspace*{-1.5cm}
%\begin{tabular}{cc}
%\includegraphics[clip=true, width=8cm]{HaL40.eps}&
%\includegraphics[clip=true, width=8cm]{HbL40.eps}\\
%\includegraphics[clip=true, width=8cm]{HaL15.eps}&
%\includegraphics[clip=true, width=8cm]{HbL15.eps}
%\end{tabular}
\includegraphics[clip=true, width=16cm]{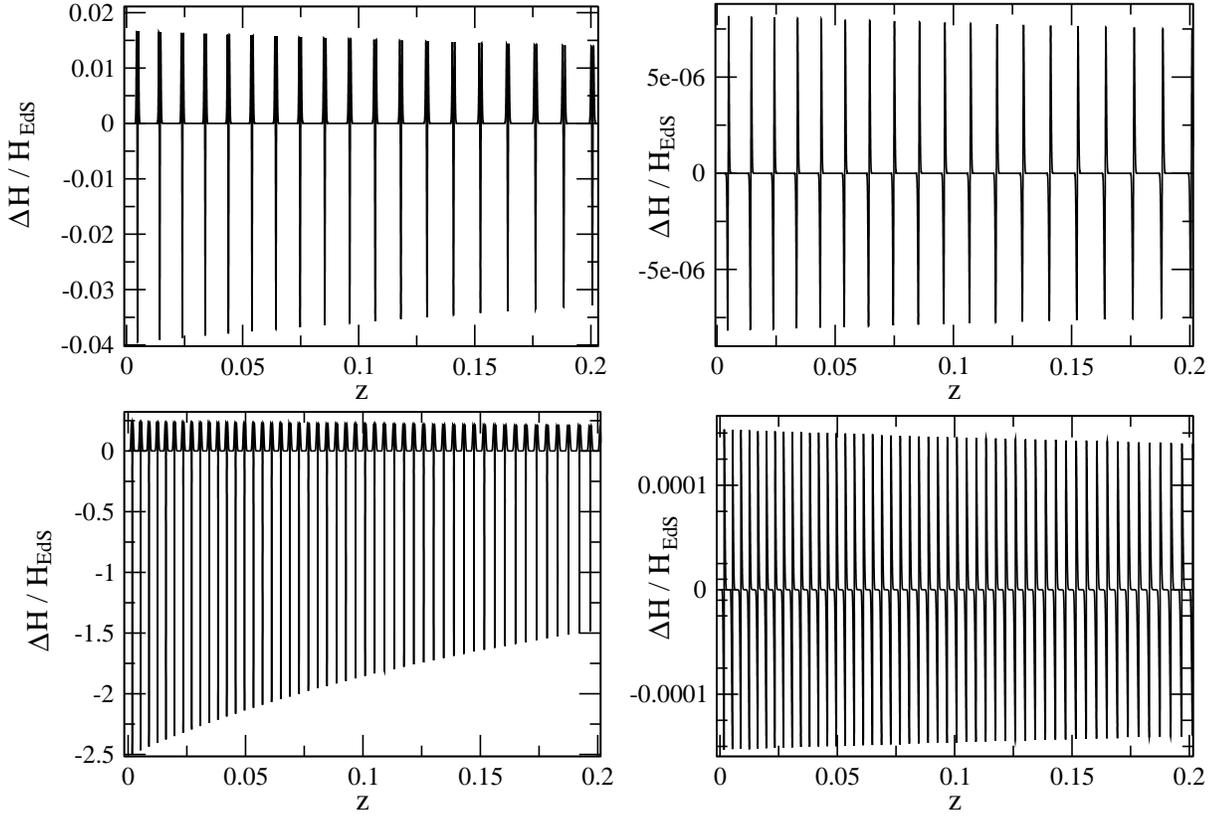}
\caption{\label{f:exp,Hubble} We show the relative differences between the expansion rates in the thin, highly concentrated over-dense wall model and the Hubble expansion in EdS universe.  The top panels are obtained with $\epsilon =10^{-9}$, $\lambda= 40 $ Mpc and $\sigma=1$ Mpc, while the bottom ones with $\epsilon=5 \times 10^{-8}$, $\lambda= 15 $ Mpc and $\sigma=1 $ Mpc. In both cases the observer is at the center of the void. The left panels show the expansion rates in the perturbed direction, while the right ones in the y-direction. The results for the cosine profile not shown here
are similar to the two top panels.}
\end{figure}

\subsection{Mimicking dark energy}\label{s:mimi}
\begin{figure}[h!]
\centering  ~  
\includegraphics[clip=true, width=12cm]{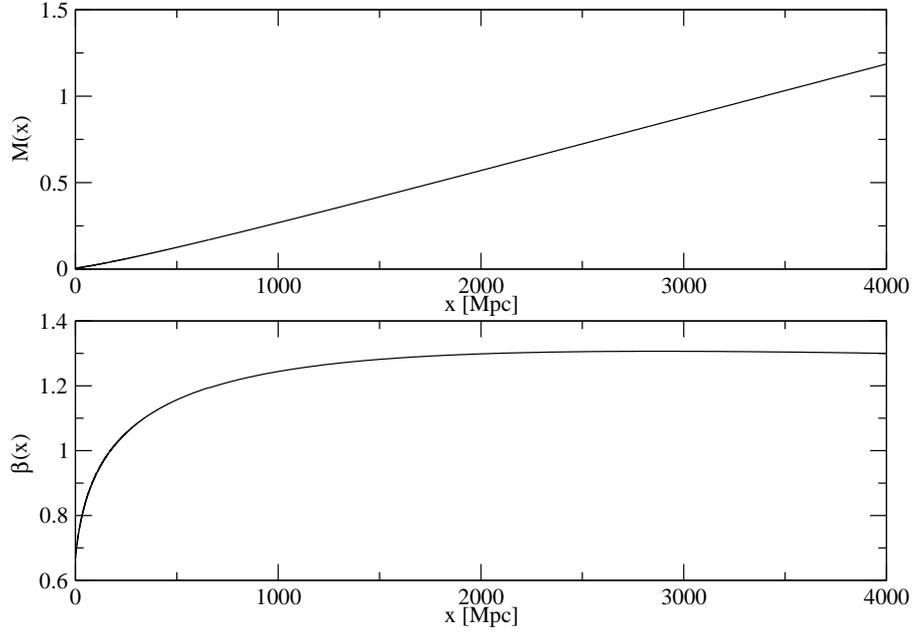}
\caption{\label{f:mimi} We show the function $M(x)$, top panel, and its derivative $\beta(x)$, bottom panel. 
In principle, there is a entire family of functions $M(x)$ parametrized by the initial value $M(0)=M_{0}$ 
that we are free to choose (appendix A). Here, we present the solutions corresponding to $ H_0 = \left. \frac{\dot a}{a}\right|_{s_{0}} = \left. \frac{\dot b}{b}\right|_{s_{0}} $.}
\end{figure}

Yoo et al.~\cite{yoo} have shown that in an LTB model every given distance--redshift
relation can be mimicked by a suitable choice of the density profile. The same is 
true for a wall universe. For a given function $D(z)$ we can find a density profile which
leads to exactly this distance--redshift relation for a photon coming in $x$-direction. 
First of all, for such a photon the shear vanishes for symmetry reasons and $\RR$ is 
given by (\ref{e:Rab}). To find the density profile, which is equivalent to finding $M(x)$
or $M(z)\equiv M(x(z))$ we have to solve the following
coupled system of six ordinary differential equations (in principle none of the other equations couples to 
(\ref{e:s2}) since both, $F_M$ and $F_\beta$ do not depend on $x$ explicitly), which is very similar to the system
solved in Ref.~\cite{yoo}:
\begin{eqnarray}
\frac{d M}{ds} &=& F_M \left( t,z, M, \beta, \zeta \right),  \label{e:s1}\\
\frac{d \beta}{ds} &=& F_\beta  \left( t,z, M, \beta, \zeta \right), \\
\frac{d x}{ds}  &=& \frac{F_M  \left( t,z, M, \beta, \zeta \right)}{\beta},  \label{e:s2} \\
\frac{d t}{ds}  &=& 1+z ,\label{e:s3} \\
\frac{d z}{ds} & =& \frac{\zeta}{\frac{dD}{dz}}, \label{24}\\
\frac{d \zeta}{ds}  &=&  - 4 \pi \left( 1 + z \right)^2 \rho D, \label{e:s6}
\end{eqnarray}
where we have defined
\bea \label{26}
\zeta &=& \frac{d z}{ds} \frac{dD}{dz} \  \  \text{and} \  \  
\beta = \frac{\frac{d M}{ds} }{\frac{d x}{ds}} = M' = \frac{2}{3t_0^2}E \, .
\eea
In Appendix \ref{app:A} we give the derivation of this system and the detailed expressions 
for $F_M$ and $F_\beta$. There, we also explain the method used to specify the initial conditions 
at the observer. All the constraints are fixed by requiring the system to have no critical points. 
Note also that $z(s)$ need not to be monotonic. If $dz/ds=0$ at a value of $s$ where 
$\zeta=dD/ds\neq 0$, the derivative $dD/dz$ is not well defined. This is, however, not the case of a $\La$CDM
Universe which we want to mimic here.   
We are then left with one initial condition, which we choose by requiring 
\bea
H_0 = \left. \frac{\dot a}{a}\right|_{s_{0}} = \left. \frac{\dot b}{b}\right|_{s_{0}},
\eea  
i.e. the value of the Hubble rate at the observer today does not depend on direction.
In Fig.~\ref{f:mimi} we show $M(x)$ as well as its derivative with respect to the 
$x$ coordinate, $\beta\left( x \right) $, for the solution mimicking the 
$\La$CDM expression for $D(z)$, for $\Om_K=0$, $\Om_m=0.3$ and $\Om_{DE}(z)
= 0.7 = $constant.
\bea \label{e:LCDM}
D(z) &=& \frac{1}{1+z}\chi_K\left(\int_0^z \frac{dz'}{H(z')}\right) \quad \mbox{ where } \\
\chi_K(r) &=&  \frac{1}{\sqrt{K}}\sin(r\sqrt{K}) \,, ~  \mbox{ and} \nonumber\\  
H(z) &=& H_0\Big(\Om_m(1+z)^3 +\Om_K(1+z)^2 +   \Om_r(1+z)^4 +\Om_{DE}(z) \Big)^{1/2}  \nonumber \,.
\eea

\begin{figure}[h!]
\centering % \vspace*{2.0cm}~  
\includegraphics[clip=true,width=12cm]{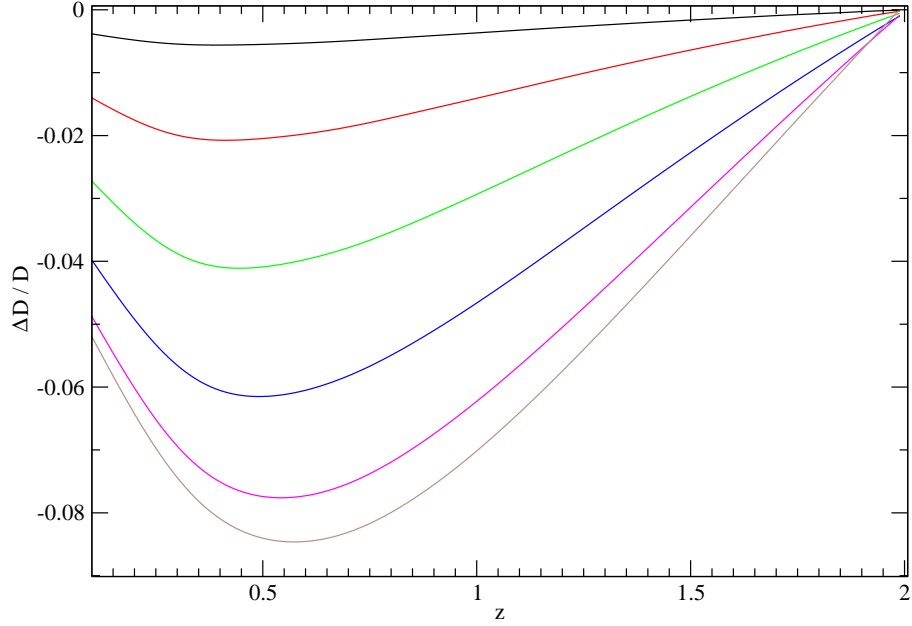}
\caption{\label{f:angles} We show the relative differences between luminosity distances for photons traveling in the $x$-direction (perpendicular to the walls) and photons observed with an angle $\theta_{0}$ (see Eq.(\ref{e:theta})). From the top to the bottom, we respectively have $\theta_{0}= 75,\,60,
\, 45, \, 30,\, 15, \, 5 $ degrees.}
\end{figure}

In Fig.~\ref{f:angles}, we show how the luminosity distance deviates when the observer looks at photons coming in with different angles $\theta_{0}$.
For $\theta_{0}=90$ degrees, we have photons traveling in $x$-direction, in this case the luminosity distance is
fitted to the one of $\La$CDM by solving 
the system of Eqs. (\ref{e:s1}-\ref{e:s6}) with the functions $M(x)$ and $\beta(x)$ shown in Fig.~\ref{f:mimi}. It is interesting to remark
that a given angle of $\theta_{0} \in\left[0;90\right]$ degrees at the observer corresponds to an angle at the emission $\theta_{e} > \theta_{0}$. 
This is a consequence of the spacetime geometry induced by the walls: due to the clustering in direction $x$, 
corresponding to $\theta=90^o$, its expansion slows down in time.

\begin{figure}[h!]
\centering  ~  
%\vspace*{0.7cm}
\includegraphics[clip=true, width=12cm]{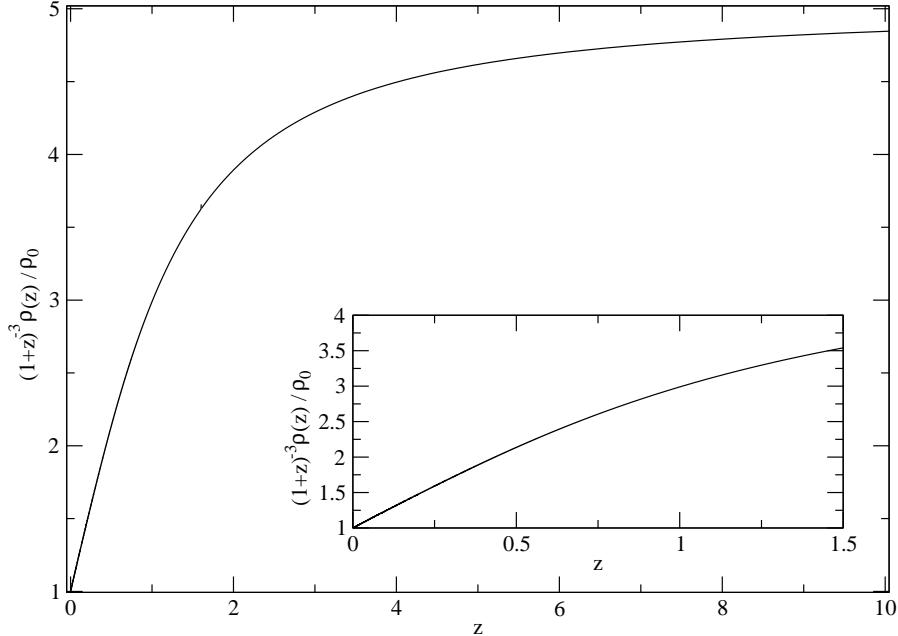}
\caption{\label{f:rho} We show the ratio of our density profile to the Einstein-de Sitter one as a function of the cosmological redshift.}
\end{figure}
\begin{figure}[h!]
\centering  ~  
\includegraphics[clip=true, width=12cm]{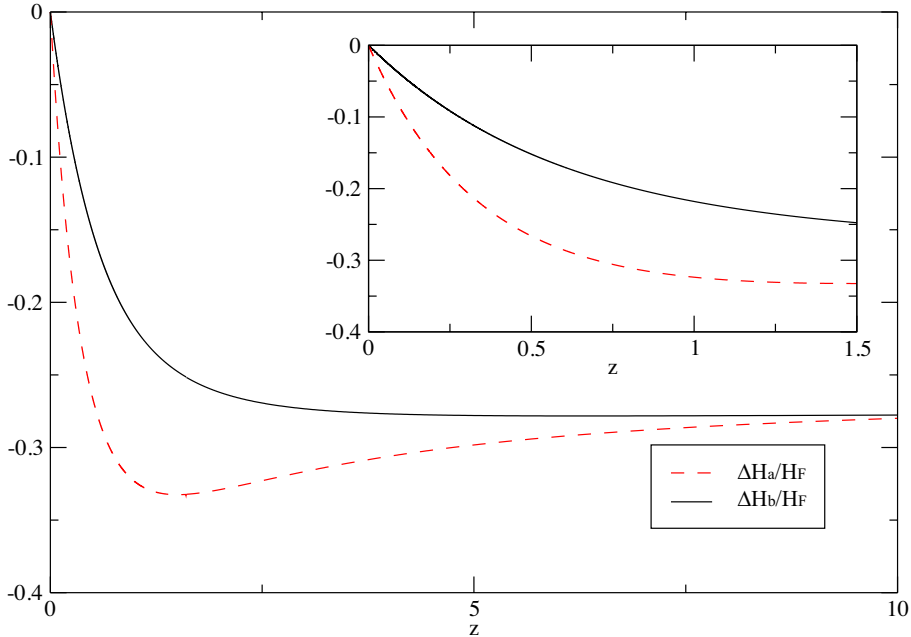}
\caption{\label{f:rates} We show the relative expansion rates in the transverse and longitudinal directions as functions of
the cosmological redshift. We use the following notation : $\Delta H_{a,b} = H_{a,b}-H_{F}$, where $H_{F}$ is the 
expansion rate in an Einstein-de Sitter universe, and $H_{a,b}$ are the expansion rates in the longitudinal and transverse
directions, normalized to the their values at the observer.}
\end{figure}
In Fig.~\ref{f:rho}, we present the density profile corrected by the isotropic expansion rate, $(1+z)^{-3} \rho(z) / \rho_{0}$, $\rho_{0}=\rho(z=0)$,  obtained for
our model to mimic $\Lambda$CDM luminosity distance. Finally, in Fig.~\ref{f:rates}, we plot the expansion rates in the longitudinal and
transverse directions, $H_{a} =\dot a / a $ and $H_{b} = \dot b / b $.
It is interesting to estimate roughly the features of the under density needed to fit $\Lambda$CDM luminosity distance.
For example, if one considers the highest redshift for which we have data from supernovae, at around $z\sim 1.7$.
This roughly corresponds to a size  $ \sim H_{0}^{-1} $.  (Of course we have another data point from the CMB. The angular size of the acoustic oscillations provides an excellent measure of the angular diameter distance to the last scattering surface, $z\simeq 1090$. But this is not very relevant in our context as the Universe is to a good approximation matter dominated from $z=2$ to $z=1090$.)  An under density of
the size of the order of the Hubble distance is necessary to mimic $\La$CDM with our walls. 
Moreover, we can also determine the ratio of the energy density normalized at the observer to the
energy density in an Einstein-de Sitter model at $z\sim 1.7$ which is about $4$. At high redshift, $z \gtrsim 10$ the
anisotropy is very small and the Universe is close to a Friedmann Universe with about 5 times the
matter density obtained from local estimates.
 
\subsection{Redshift drift}\label{s:redr}
\begin{figure}[h!]
\centering  ~  
\vspace{1cm}\\ 
\includegraphics[clip=true, width=12cm]{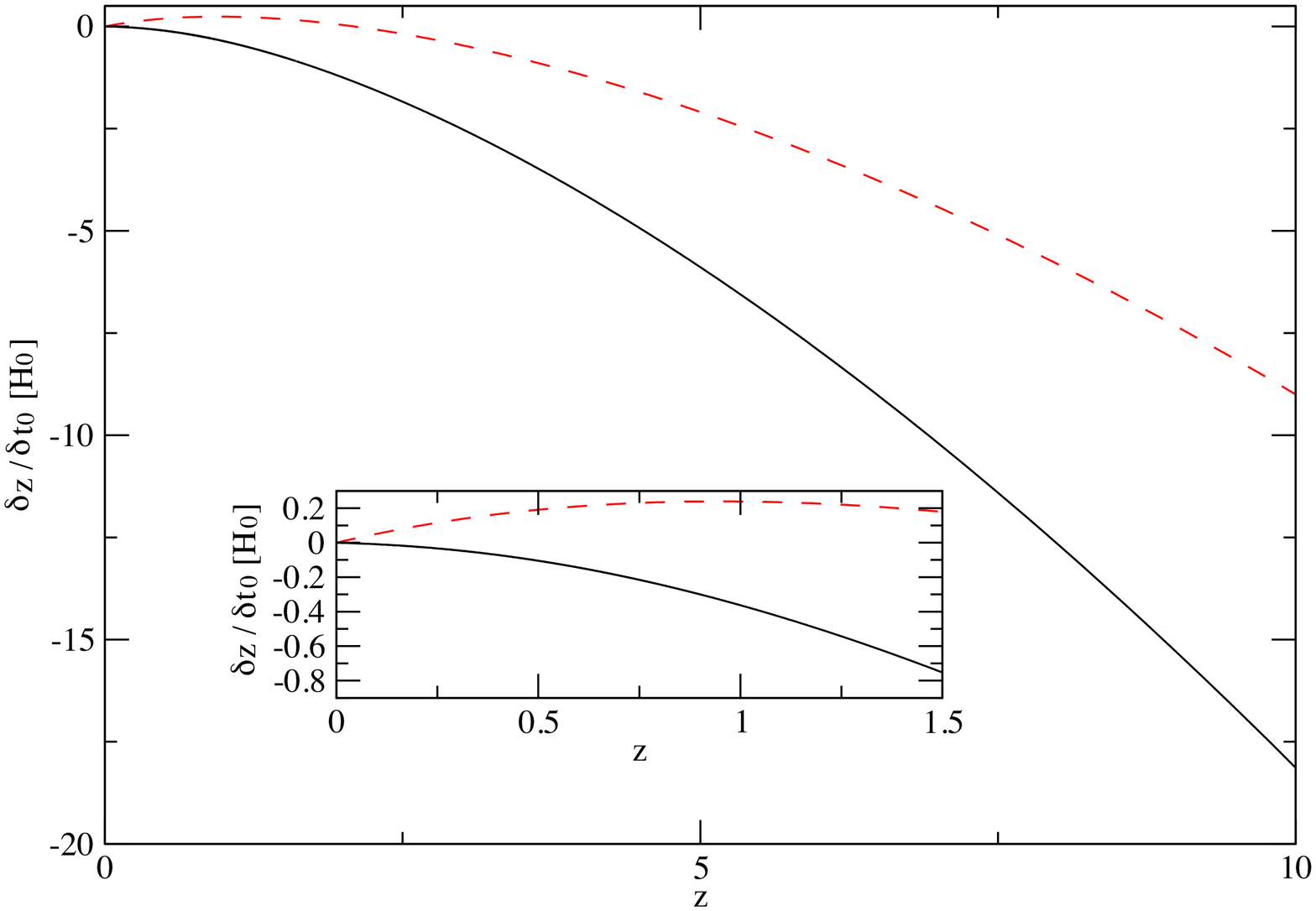}
\caption{\label{f:rdrift} We show the redshift drift for the wall Universe (black solid line) which mimics the distance
redshift relation of $\La$CDM and compare it with the redshift drift of the latter  (red dashed line).}
\end{figure}

In the previous section we have fixed $M (x)$ to reproduce the distance redshift relation of $\La$CDM universe. 
Of course, having one free function to play with, namely $M(x)$, we expect to be able to
fit one function, in our case $D(z)$. If we now proceed to another, independent observable,
we shall most probably not fit it. We have done this by looking at the redshift drift, defined as the rate of change of the redshift of a co-moving source per unit of observer time. In a 
Friedmann Universe the redshift drift is simply
\be\label{e:reddrift}
\frac{dz}{dt_0} \equiv \lim_{\De t_0 \ra 0}\frac{z(t_s + \De t_s)-z(t_s)}{\De t_0} = H_0(1+z) -H(z) \,,
\ee 
where $H(z)=H(t_s)$ and $H_0$ denote the Hubble parameter at the source position at time $t_s$ and at the 
observer at the moment $t_0$. We have computed the corresponding function (for light rays in $x$-direction)
from our solution $M(x)$. The general expression for the redshift drift of a wall Universe in $x$--direction is
 (see Appendix \ref{app:drift}),
\begin{eqnarray} \label{reddrift1}
\frac{dz}{dt_0} &=& \left( 1+z \right) \int_0^z \left( \frac{\ddot b'}{\dot b'}\right)\left( 1 + z' \right)^{-2} dz'  \nonumber \\
&=&   - \left( 1+z \right) \int_0^z \left( 4 \pi G \rho - \frac{2 M}{b^3} \right) \frac{a}{\dot a}\left( 1 + z' \right)^{-2} dz' \ .
\end{eqnarray}
Since we do not require $M_0 =0$ as in LTB model, we can in principle have a positive redshift 
drift at low redshift; but we do not obtain  this for our best fit profile $M(x)$ with $t_B(x)\equiv 0$.
The result is compared with $\La$CDM in Fig.~\ref{f:rdrift}.

Clearly the redshift drift for the two cosmologies are very different. We do have a second
function to play with, the bang time $t_B(x)$,  so that we could probably fix this observable.
This has been done for LTB models in~\cite{pedro}. However, as it is shown there, models
which have both, the same redshift distance relation and the same redshift drift as $\Lambda$CDM can be ruled
out with a third observable, the kinematic Sunyaev-Zel'dovich effect which comes from
the recession velocity of clusters.

\section{Conclusions}\label{s:con}

We have studied the effect of matter perturbations on the luminosity distance in a model with
planar symmetry described by the metric (\ref{e:metric}). 
Considering 'realistic' walls we find that the effect from density inhomogeneities is very small, it nearly 
averages out. It leads to fluctuations of the luminosity distance around the 'background' distance, but not to a significant global shift. Our results (Fig.~\ref{f:exp}) show that these fluctuations are due to matter inhomogeneities at the source and the observer positions, without any relevant contribution from the integrated effects of light propagation, like in the linear approach (\ref{linearregime}). Hence we can not mimic acceleration with many
dense walls which grow by gravitational instability. Since we consider pressure-less matter only, the amplitude of density fluctuations is limited by the presence of singularities. This is a limitation of the model.

After having shown that 'realistic' wall models can not reproduce the observed distance-redshift relation, we have determined the density profile which can mimic it. We have fixed the free function of our model, $M(x)$, to mimic the luminosity (or angular) distance of the $\Lambda$CDM universe. We have shown that the observation of the redshift drift can distinguish between this model and $\Lambda$CDM. Abandoning the assumption of
an uniform bang time we could arrange  the second degree of freedom, $t_B(x)$, to fit the redshift drift too. We have found that the redshift drift in our model can be positive at low redshift, contrarily to the LTB model \cite{yoodrift}.

With our solution $M(x)$ we can fit $\Lambda$CDM distance for photons coming in $x$-direction for positive $x$ only. This preferred direction corresponds to the radial incoming direction for LTB model. The deviation from 
$\Lambda$CDM for photons coming from different  angles is typically a few percent (see Fig. \ref{f:angles}).

\section*{Acknowledgements}
This work is supported by the Swiss National Science Foundation.

\appendix
\section{Derivation of the system of differential equations and initial conditions}
\label{app:A}
\subsection{The system}
Here we derive in more detail the system (\ref{e:s1}) to  (\ref{e:s6}) and give the 
initial conditions used for the solution.

Since we choose the photon affine parameter such that $n^0|_0 =1$ we have
$$1+z(s) = n^0(s) = \frac{dt}{ds} .$$
Furthermore, the null condition for a light ray in $x$--direction implies
\begin{equation}\label{16}
\left(\frac{dt}{ds}\right)^2 = \left( \frac{b'}{E}\right)^2 \left(\frac{d x}{ds}\right)^2 =
 \left( \frac{2 b'}{3 t_0^2 M'} \right)^2\left(\frac{d x}{ds}\right)^2 .
\end{equation}
The geodesic equation gives
\be\label{e:geo} 
\frac{dz}{ds} = \frac{d^2t}{ds^2} =-\frac{\dot a}{a} (1+z)^2 
  = - \frac{\dot b'}{b'} (1+z)^2  \, . 
  \ee
Hence,  when the expansion in $x$-direction changes into contraction, $\dot a =0$, also $dz/ds$ passes through zero.
However, this does not happen in our case which mimics $\La$CDM.
Noting that geodesics in x-direction have no shear, the Sachs focusing equation yields
\begin{equation}
\frac{d^2z}{ds^2} \frac{dD}{dz} + \left(\frac{dz}{ds}\right)^2 \frac{d^2 D}{dz^2} = 
- 4 \pi G \rho \left( 1 + z \right)^2 D,
\end{equation}
where we have used $\RR= 4 \pi G \left( 1 + z \right)^2 \rho$.
We can now rewrite these equations in terms of the system (\ref{e:s1}) to  (\ref{e:s6}). To find the
functions $F_M$ and  $F_\beta$ we first derive the following useful relations
\bea
\dot\tau &=& \frac{2}{3t} \tau , \\
\tau' &=& \tau \left( 2 \frac{\frac{d \beta}{ds}}{\frac{dM}{ds}} - \frac{2}{3} \frac{\beta}{M} \right) , \\
b' &=& - \frac{8}{9t_0^4}\frac{\tau}{\beta}\left( S-2\tau S' \right) - \frac{16}{3t_0^4} \tau^2 \frac{M}{\beta^2} \frac{\frac{d \beta}{ds}}{\frac{dM}{ds}}S' , \\
\dot b' &=& - \frac{16}{27t_0^4} \frac{\tau}{t} \frac{1}{\beta}  \left( S- 3 \tau S' - 2 \tau^2 S'' \right)   - \frac{32}{9t_0^4} \frac{M}{\beta^2} \frac{\frac{d \beta}{ds}}{\frac{dM}{ds}}  
\frac{\tau^2}{t} \left( 2 S' + \tau S'' \right) .
\eea
Here $S'$ always indicates the derivative of $S$ with respect to its argument $\tau$ while as
for all other functions of $(t,x)$ the prime denotes the partial derivative w.r.t. $x$ and the dot 
the one w.r.t. $t$.
The null condition for the light ray can be written as
$$ \frac{dM}{ds} A_1 + \frac{d\beta}{ds} B_1= \pm 1 ,$$
with
\begin{eqnarray}
A_1&=&- \frac{16}{27 t_O^6} \frac{\tau}{\beta^3 \left( 1 + z \right) } \left( S - 2 \tau S' \right) ,
 \\
B_1&=&- \frac{32}{9 t_O^6}  \frac{\tau^2 M}{ \beta^4\left( 1 + z  \right)}S' . 
\end{eqnarray}
The geodesic equation takes the form
\begin{equation}
\frac{dM}{ds} A_2 +\frac{d\beta}{ds} B_2=0,
\end{equation}
where 
\begin{eqnarray}
A_2&=&- \frac{\zeta}{\frac{dD}{dz}} \frac{8}{9t_0^4} \frac{\tau }{\beta} \left( S- 2\tau S' \right)   - \left( 1 + z \right)^2 \frac{16}{27t_0^4} 
\frac{\tau}{t} \frac{1}{\beta} \left( S - 3 \tau S' - 
    2 \tau^2 S'' \right) , \\
B_2 &=& - \frac{\zeta}{\frac{dD}{dz}} \frac{16}{3t_O^4} \tau^2 \frac{M}{\beta^2} S' 
- \left( 1 + z \right)^2 \frac{32}{9 t_0^4} \frac{M}{\beta^2} \frac{\tau^2}{t} \left( 2 S' + \tau S'' \right),
\end{eqnarray}
with $\zeta = \frac{dD}{ds} = \frac{dz}{ds}\frac{dD}{dz}$. From this we infer
\begin{eqnarray}
F_M \left( t,z,M, \beta, \zeta \right)&=&\pm \frac{B_2}{A_1B_2-A_2B_1},    \\
F_\beta \left( t,z,M,\beta, \zeta \right)&=&\mp\frac{A_2}{A_1B_2-A_2B_1}.
\end{eqnarray}
Since $\tau$ is a function of $M$, $\beta$ and $t$, we now have expressed everything
in terms of our variables $(t,z,M,\beta, \zeta)$ and the given function $D(z)$. Explicitly, 
$F_M$ and $F_\beta$ are given by
\begin{eqnarray}
F_M&=&\pm  \frac{3 t_O^2}{4} \beta \left( \frac{6M}{t} \right)^{2/3}  \frac{\frac{\zeta}{\frac{dD}{dz}} \frac{3t}{2}\frac{S'}{1+z} + \left( 1 + z \right) \left( 2 S' + \tau S'' \right) }{  S S'  + \tau S S'' - \tau S'^2  }\ ,  \\
F_\beta &=& \pm \frac{1}{18t_O^2} \frac{1}{M} \left( \frac{6M}{t} \right)^{4/3}  \frac{ \frac{\zeta}{\frac{dD}{dz}}  \frac{3t}{2} \frac{\left( S - 2 \tau S' \right)}{1+z} + \left( 1 + z \right)  \left( S-\tau S' -  2 \tau^2 S'' \right)}{  S S'  + \tau S S'' - \tau S'^2  }\ . 
\end{eqnarray} 

\subsection{Initial conditions}
Let us now turn to the initial conditions at $s_0=0$. Without loss of generality we
can set $x(0)=0$.  Clearly also $z(0)=0$. From definition (\ref{26}) we have 
\be
\zeta \left( 0 \right) = \left. \frac{dD}{ds} \right|_{s=0}.
\ee
Since this is an initial condition for the Sachs focusing equation, we have consistently with our affine parameter normalization~\cite{SPE, Perlick},
\be \label{app:a18}
\zeta \left( 0 \right) =-1.
\ee
From (\ref{24}) we note that our system of coupled differential equations has a critical point $z_{cr}$ defined by 
\be\label{app:a19}
\left. \frac{d D}{dz} \right|_{z=z_{cr}} =0.
\ee
For our $\Lambda$CDM parameters $z_{cr} \approx 1.6$. 
To obtain a regular solution we must therefore impose $\zeta\left( z_{cr} \right) =0$.
We remark that Eqs.~(\ref{app:a18}) and (\ref{e:geo}) imply
\be\label{app:a20}
\frac{\dot a}{a} = H_0,
\ee
where we have used 
\bean\label{app:a21}
\left. \frac{dD}{dz} \right|_{z=0} = H_0^{-1}.
\eean
Hence the rate expansion in $x$-direction coincides with the measured Hubble expansion. In order to solve the system of five
differential equations (Eq.~(\ref{e:s2}) is an independent equation, since the solution $x\left(s\right)$ can 
also be inferred from Eq.~(\ref{e:s3}) 
via the null condition), five initial 
conditions are needed. However, we only have two of them
\bea\label{app:a22}
z\left(0 \right) = 0 \qquad  \zeta \left( 0 \right) = -1.
\eea
 We have two other constraints which we must satisfy at the critical
point where (\ref{app:a19}) holds. Denoting the affine parameter at the critical point by $s_{cr}$, we have
\bea\label{app:a23}
z\left( s_{cr}\right) = z_{cr} \qquad  \zeta \left( s_{cr} \right) = 0.
\eea
These lead to two other initial conditions which can be determined using the shooting method.
One remaining constraint is needed and we fix it  by requiring 
\be\label{app:a24}
\left. \frac{\dot a}{a}\right|_{0} =\left. \frac{\dot b}{b}\right|_{0} = H_0.
\ee
This last condition fixes $M(0)$ and makes sure that the Hubble rate measured today is the same in any direction. 
We then numerically integrate the system from the critical point to the observer by varying
the three remaining conditions at the critical point until the initial conditions (\ref{app:a22}) and (\ref{app:a24}) are satisfied. This matching is obtained by using
the three dimensional Newton-Raphson method. Once the desired precision has been reached, the two remaining initial conditions $\beta(0)$ and $t(0)$ can simply be read from the numerical data.

\section{Derivation of the system of differential equations for the redshift drift}\label{app:drift}
The redshift drift for a LTB model has been derived in \cite{yoodrift}. This approach can also be applied 
to our model. The null condition for the light ray (in $x$-direction) and the geodesic equation lead to
\begin{eqnarray}
\frac{dz}{dx} = \frac{\dot b'}{E} \left( 1 + z \right), \qquad
\frac{dt}{dx} = -\frac{b'}{E}.\label{app:b1}
\end{eqnarray}
We consider two infinitesimally close geodesics at fixed comoving position $x$, parametrized by
\bean
\left\{ z_c, t_c \right\} \   \   \text{and}  \  \  \left\{ z_c + \delta z, t_c + \delta z\right\}.
\eean
Since the geodesic $\left\{ z_c, t_c \right\}$ satisfies (\ref{app:b1}), it follows
\begin{eqnarray*}
\frac{ d \delta z}{dx} &=& \frac{\ddot b' }{E} \left(  1 + z \right) \delta t + \frac{\dot b'}{E} \delta z, \\
\frac{d \delta t}{dx} &=& - \frac{\dot b'}{E} \delta t.
\end{eqnarray*}
Then, inserting (\ref{app:b1}) we obtain
\begin{eqnarray}
\frac{d \delta z}{dz} &=& \frac{\ddot b'}{\dot b'} \delta t + \frac{\delta z}{1+z}, \label{app:b3} \\
\frac{d \delta t}{dz} &=& - \frac{\delta t}{1+z}. \label{app:b4}
\end{eqnarray}
Integrating (\ref{app:b4}) we find
\bean
\delta t = \frac{\delta t_0}{1+z}.
\eean
This solution together with (\ref{app:b3}) leads to
\bean
\frac{d}{dz} \left( \frac{\delta z}{\delta t_0}\right)=\frac{1}{1+z} \left( \frac{\ddot  b'}{\dot b'} + \frac{\delta z}{\delta t_0} \right).
\eean
This equation is solved by (\ref{reddrift1}).
Deriving the Einstein equation (\ref{e:00}) twice (once w.r.t. $x$ and once w.r.t. $t$),  we obtain 
\be
\ddot b' = \frac{2M b'}{b^3}- \frac{M'}{b^2}.
\ee
 With (\ref{e:0i}) and (\ref{e:M'}) this results in the second line of (\ref{reddrift1}).

\section{The linearized approach}\label{app:linear}
We determine the luminosity distance within linear perturbation theory for small deviations from a 
Friedmann--Lema\^\i tre background. Let us define
\begin{eqnarray}
a(t,x) &=& \bar{a}(t) \left( 1 + \epsilon f (t,x) \right), \\
b(t,x) &=& \bar{a}(t) \left( 1 + \epsilon g(t,x) \right), \\
\rho(t,x)&=& \bar{\rho} (t) \left( 1 + \epsilon \delta (t,x) \right) ,
\end{eqnarray}
where the unperturbed quantities $\bar a \left( t \right)$, $\bar \rho \left( t\right)$ satisfy the Einstein 
equations for a flat matter dominated Friedmann universe (EdS). The perturbed quantities are determined 
by  the Einstein equations at first order in $\ep$,
\begin{eqnarray}
\frac{-6t_0^{4/3} g'' + 4 t^{1/3} \left( \dot f + 2 \dot g \right)}{3t^{4/3}}&=& 8\pi G \bar{\rho} \delta,  \label{C4}\\
\dot g' &=&0 , \\
t^{1/3} \left( 2 \dot g + t \ddot g \right)& =&0, \\
t_0^{4/3} g'' - t^{1/3} \left( 2 \dot f + 2 \dot g + t \left( \ddot f + \ddot g \right) \right)&=&0. \label{C7}
\end{eqnarray}
Neglecting the decaying mode and imposing that at the beginning the scale factors in all three  
directions agree, we obtain~\cite{collins},
\begin{eqnarray} \label{c8}
g&=&\frac{\delta_O}{3}, \\
f&=& \frac{3}{10} \delta_O'' t_O^{4/3} t^{2/3} + \frac{\delta_O}{3}, \label{c9}
\end{eqnarray}
where $\delta_O \left( x \right) = \delta \left( t, x \right) + f\left( t, x \right) + 2 g\left( t, x \right)$ is independent of time.
This is a consequence of energy conservation and can also be derived by combining (\ref{C4}) to (\ref{C7}).

We are interested in finding the relation between $\delta_O$ and $M, E$ in the perturbative regime. Following \cite{zakharov} we expand the solution (\ref{e:En0Mp}, \ref{e:En0Mpt}) around $\eta =0$ in terms of $\frac{t_B\left( x\right)}{t} \ll 1$ and $ \frac{E^3 t }{M} \ll 1$. Comparing the expanded solution with the linear one we find
\begin{equation}
M= \frac{2}{9t_0^2} \left( 1 + \epsilon \delta_O\right), \  \   \  E= \frac{\epsilon \delta_O'  }{3}.
\end{equation}
With the ansatz (\ref{e:Mh}, \ref{e:Eh}) we can identify $\delta_O \left( x \right)$ with $h \left( x \right)$ in the perturbative regime.

The angular distance is determined by Sachs focusing equation (\ref{e:foc}). We note that the shear term does not contribute to first order. Since light propagation is not affected by a conformal transformation, it is convenient to work with the conformally related geometry
\be
ds^2 = - d \eta ^2 + \left( 1 + 2 \epsilon f \right)dx^2+ \left( 1 + 2 \epsilon g \right) \left( dy_1^2 + dy_2^2 \right) \,.
\ee
From this, we compute the Christoffel symbols (here we denote the derivative w.r.t. the conformal time $\eta$
by a dot)
\begin{eqnarray*}
\begin{array}{ccccccc}
&  &  &  \Gamma^0_{11} \cong \epsilon \dot f, & &   &  \Gamma^0_{22} =\Gamma^0_{33} \cong \epsilon \dot g, \\
\Gamma^1_{10} \cong \epsilon \dot f, &  &  &   \Gamma^1_{11} \cong \epsilon f',  &  &  &\Gamma^1_{22} =\Gamma^1_{33} \cong- \epsilon g', \\ 
&  &  &\Gamma^2_{20}= \Gamma^3_{30} \cong \epsilon \dot g,  &    &  &  \Gamma^2_{21} = \Gamma^3_{31} \cong \epsilon g' ,
\end{array}
\end{eqnarray*}
and the Ricci tensor
\begin{eqnarray*}
R_{00} &\cong& - \epsilon \left( \ddot f + 2 \ddot g \right), \\
R_{10}&\cong& -2 \epsilon \dot g' , \\
R_{11} &\cong& \epsilon  \left( \ddot f - 2 g'' \right), \\
R_{22}= R_{33} & \cong & \epsilon \left( \ddot g - g'' \right).
\end{eqnarray*}
At 0-order we are free to parametrize the affine parameter $s$ such that $\bar n^0=1$ and $\bar n^i = \delta^{i1}$ (we are interested in the distance in $x$-direction). With this we obtain the coefficient $\RR$ 
\bean
\RR= - \epsilon \left( \ddot g + g'' + 2 \dot g' \right).
\eean
Consistently with the parametrization of the affine parameter $s$ such that $n^0 \left( s_0\right) =1$, the initial conditions are $D\left( s_o \right) =0$ and $D'\left( s_o \right) =-1$. After an integration by parts we find 
the solution to Sachs focusing equation (\ref{e:foc}),
\be \label{c12}
D\left( s\right) = \left( s_o -s \right) \left( 1 + \epsilon g \left( s_o \right)+ \epsilon g\left( s\right) \right) + 2 \int_{s_o}^s ds' \epsilon g\left( s' \right).
\ee
With the above initial conditions for the Sachs focusing equation, we consider a thin light bundle with the vertex at the observer position. Hence the solution (\ref{c12}) is the angular diameter distance, see~\cite{SPE}. To determine the luminosity distance we have to compute also the redshift, using the geodesic equation for $n^0$,
\be
1 + z = \frac{\left. g_{\mu \nu} n^\mu u^\nu \right|_e}{\left. g_{\mu \nu} n^\mu u^\nu \right|_o}= \left. n^0 \right|_e = 1 - \int_{s_o}^{s_e} ds \  \epsilon \dot f,
\ee
where $|_e$ denotes the emission point, the source, and we denote the affine parameter at the source by $s_e$. With the same geodesic equation we derive the relation between the conformal time $\eta$ and the affine parameter $s$, $n^0 = d\eta/ds$,
\be
\eta_o- \eta_e = s_o - s_e + \int_{s_o}^{s_e} ds \int_{s_o}^{s} ds' \epsilon \dot f(s'). 
\ee
In terms of conformal time the luminosity distance then becomes
\be
D_L\left( \eta_e \right)  = \left( \eta_o - \eta_e \right) \left( 1 + \epsilon g_o + \epsilon g_e - 2 \int_{\eta_o}^{\eta_e} d\eta \ \epsilon \dot f \right) 
 + 2 \int_{\eta_o}^{\eta_e} d\eta \ \epsilon g -  \int_{\eta_o}^{\eta_e} d\eta \int_{\eta_o}^\eta d\eta' \epsilon \dot f. 
\ee
All of this is valid in the conformal geometry, where the expansion of the Universe is divided out. 
Taking into account the expansion of the universe, changes the relation between the affine parameter and conformal time. The luminosity distance scales as~\cite{CRA}
\bean
\tilde D_L = \frac{\bar a^2 \left( \eta_o \right)}{\bar a \left( \eta_e \right)} D_L  = \frac{D_L}{\bar a \left( \eta_e \right)} = \left( 1 + \bar z_e \right) D_L.
\eean
Since conformal time is not an observable quantity, we rewrite the distance in term of the observed redshift. We define the observed redshift as $z_e= \bar z_e + \delta z_e$ and we compute the correction term. 
The same calculation as presented in Ref.~\cite{CRA} leads to
\be \label{c16}
\left( \frac{d}{dz} \tilde D_L \right) \delta z_e =\left(  \left( \eta_o - \eta_e \right) +  \mathcal{H}_e^{-1} \right) \delta z_e,
\ee
where
\be
\delta z_e = -\left( 1  + z_e \right) \int_{\eta_o}^{\eta_e} d\eta \ \epsilon \dot f.
\ee 
Subtracting (\ref{c16}) we obtain the distance--redshift relation
\begin{eqnarray}
\tilde D_L\left( z_e\right) &= &\left( 1 +z_e \right)  \left( \eta_o - \eta_e \right) \left( 1 + \epsilon g_o + \epsilon g_e  - \int_{\eta_o}^{\eta_e} d\eta \ \epsilon \dot f \right)  \\
&+& \left( 1 +z_e \right) \left(   2 \int_{\eta_o}^{\eta_e} d\eta \ \epsilon g -  \int_{\eta_o}^{\eta_e} d\eta \int_{\eta_o}^\eta d\eta' \epsilon \dot f \right) 
+ \frac{1+z_e}{\mathcal{H}_e}  \int_{\eta_o}^{\eta_e} d\eta \ \epsilon \dot f. \nonumber
\end{eqnarray}
With
\bean
 -  \int_{\eta_o}^{\eta_e} d\eta \int_{\eta_o}^\eta d\eta' \epsilon \dot f =\left(\eta_e - \eta_o \right) \int_{\eta_e}^{\eta_o} d\eta \ \epsilon \dot f 
  + \int_{\eta_e}^{\eta_o} d\eta \int_{\eta_e}^\eta d\eta' \epsilon \dot f,
\eean
we can rewrite the above expression in the form as
\begin{eqnarray}
\tilde D_L\left( z_e\right) &= &\left( 1 +z_e \right)  \left( \eta_o - \eta_e \right) \left( 1 + \epsilon g_o + \epsilon g_e  \right)  \\
&+& \left( 1 +z_e \right) \left(   -2 \int_{\eta_e}^{\eta_o} d\eta \ \epsilon g +  \int_{\eta_e}^{\eta_o} d\eta \int_{\eta_e}^\eta d\eta' \epsilon \dot f \right) 
- \frac{1+z_e}{\mathcal{H}_e}  \int_{\eta_e}^{\eta_o} d\eta \ \epsilon \dot f \ . \nonumber
\end{eqnarray}
Using the solutions (\ref{c8}, \ref{c9}) we express the distance in terms of $\delta_O \left( \eta \right)$. Conformal time is defined as
\bean
d \eta = \frac{dt}{\bar a \left( t \right)} \ \ \Rightarrow \ \ \eta \left( t \right) = 3 t^{1/3} t_0^{2/3}, \ \ \text{setting}\ \  \eta\left( 0\right)=0.
\eean
This leads to
\begin{eqnarray*}
g\left( \eta, x \left( \eta \right) \right) &=&\frac{\delta_O \left( x(\eta) \right) }{3}, \\
f\left( \eta, x \left( \eta \right) \right) &=& \frac{1}{30} \delta_O'' \left(  x(\eta) \right) \eta^2 + \frac{\delta_O \left(  x(\eta) \right) }{3}, \\
\dot f \left( \eta, x \left( \eta \right) \right) &=& \frac{1}{15} \delta_O'' \left( x(\eta) \right) \eta ,
\end{eqnarray*}
and consequently to the following distance--redshift relation
\begin{eqnarray}
D_L \left( z_e \right) &=&\left( 1 +  z_e \right) \left( \eta_O -\eta_e \right) \left( 1 + \frac{\epsilon}{3} \left( \delta_O \left(  x(\eta_o)  \right) + \delta_O \left(  x(\eta_e)  \right) \right)  \right)   - \left( 1 +  z_e \right)  \int_{\eta_e}^{\eta_o}    \frac{2\epsilon}{3} \delta_O \left( x(\eta) \right) d\eta\nonumber \\
&+ &  \left( 1 +  z_e \right)   \int_{\eta_e}^{\eta_o} d\eta \int_{\eta_e}^\eta d\eta' \frac{\epsilon}{15} \delta_O'' \left(  x(\eta') \right) \eta'      -   \frac{  1 +  z_e }{\mathcal{H}_e}   \int_{\eta_e}^{\eta_o} d\eta \ \frac{\epsilon}{15} \delta_O'' \left(  x(\eta) \right) \eta     .
\end{eqnarray}

\end{document}